\documentclass{osa-article}

\journal{osac}
\newcommand{\comment}[1]{}

\articletype{Research Article}
\usepackage{float}
\usepackage{booktabs}
\usepackage{xcolor}
\usepackage{ulem}
\usepackage{amsmath}

\begin{document}
\title{High Repetition-Rate Pulse Shaping of a Spectrally Broadened Yb Femtosecond Laser}

% \author{Author name(s)}
% \address{Author affiliation and full address}
% \email{e-mail address}
%%Uncomment the following line to override copyright year from the default current year.
%\copyrightyear{2022}

\author{Julia Codere\authormark{1,a}, Michael Belmonte\authormark{1,a}, Brian Kaufman\authormark{1,2}, Michael Wahl\authormark{1}, Eric Jones\authormark{1}, Martin G Cohen\authormark{1}, Thomas Weinacht\authormark{1}, and Ruaridh Forbes\authormark{2,*}}

\address{\authormark{1} Department of Physics and Astronomy, Stony Brook University, Stony Brook, New York 11794, USA\\
\authormark{2} Linac Coherent Light Source, SLAC National Accelerator Laboratory, Menlo Park, California 94025, USA\\
\authormark{A} These authors contributed equally to this work.}

\email{\authormark{*}ruforbes@slac.stanford.edu} %% email address is required

\begin{abstract}
We demonstrate compression and shaping of few cycle pulses from a high average power Ytterbium laser system. The pulses from a commercial 20~W, 100~kHz Yb laser system are spectrally broadened in two-stages using gas-filled, stretched hollow-core fibers and then compressed and shaped in an acousto-optic modulator-based pulse-shaper. The pulse-shaper allows for compression, characterization, and shaping all in one system, producing $\sim$10~fs pulses with 50~$\mu$J of energy.  
%We explore the use possibility of direct compression and manipulation of high energy and high repetition rate pulses from a ytterbium (Yb) laser after spectral broadening in a gas-filled hollow-core fiber system. 
\end{abstract}

\section{Introduction}
High-average-power, ultrafast laser systems are increasingly in demand for many experiments which benefit from high repetition rates that enable highly differential measurements, or ones that require a very low yield per laser shot (e.g. coincidence measurements) \cite{ullrich2003recoil,zhao2017coincidence,boguslavskiy2012multielectron}. While there are commercial Ytterbium laser systems available which can provide high average power with repetition rates of 100~kHz and greater, they typically produce pulses with durations greater than 100~fs. These long pulses are not suitable for many time-resolved spectroscopy or strong-field light—matter experiments, which require few cycle pulses to resolve the fastest molecular dynamics \cite{goulielmakis2008single,sandor2016strong,zherebtsov2011controlled}.  Here we report on the spectral broadening and subsequent compression of 250~fs laser pulses from a 100~kHz, 20~W commercial Ytterbium laser system (Light Conversion Pharos) using a two-stage, stretched hollow-core fiber apparatus in conjunction with an acousto-optic modulator (AOM) based ultrafast optical pulse-shaper \cite{fetterman1998ultrafast}. Our system is capable of producing arbitrarily shaped pulses as short as 10~fs, with energies between 30 and 50~$\mu$J, at repetition rates of up to 50~kHz.

Shortening the duration of an ultrafast laser pulse can be broken into two stages: (1) broadening the pulse in the frequency domain, and (2) compression of the pulse in the time domain. There are many techniques to accomplish spectral broadening, including stretched hollow core fibers, Herriott cells, thin plates and filamentation \cite{nisoli1996generation, nagy2020generation, silletti2023dispersion, lu2018sub, he2017high, hauri2004generation}. While there are various advantages and disadvantages associated with each approach \cite{nagy2021high, Gallmann2007Comparison}, they all make use of self phase modulation 
to produce the new spectral bandwidth. 
Pulse compression is then usually achieved by applying negative group-delay dispersion (GDD) to the pulses with chirped mirrors \cite{mayer1997ultrabroadband}. Few cycle pulses ($\sim$10~fs) have been generated at high average power by means of HCF-based spectral broadening and chirped mirror-based pulse compression \cite{nagy2019generation, cardin20150, gebhardt2017nonlinear}. However, using chirped mirrors and third-order dispersion-compensating material \cite{timmers2017generating} does not allow for programmable pulse shaping or direct compensation of higher order dispersion, which can be accomplished with an AOM-based, ultrafast pulse-shaper \cite{jones2019dual,catanese2021acousto}. A further advantage of the direct dispersion control using an AOM-based pulse-shaper is that it can be used for pulse characterization through dispersion scans (D-scan) \cite{miranda2012characterization} and collinear frequency resolved optical gating (CFROG) \cite{amat2004ultrashort}. 

Here, we demonstrate the production of an ultrabroadband spectrum with cascaded fibers and the ability to compress, control, and measure the pulses using an AOM-based 4f pulse-shaper. This work builds on our previous work of pulse compression techniques with a Ti:sapphire laser system \cite{catanese2021acousto}, but now uses a commercial Yb laser system. This enables us to perform pulse shaping at high repetition-rates (50~kHz) and high pulse energies ($\sim$100 $\mu$J). We are able to compress 250~fs pulses from the output of the laser to ones that are $\sim$10~fs, a compression factor of about 25. The combination of high average power and fine control of a broadened spectrum allows for highly-differential, time-resolved spectroscopy measurements.

\section{Apparatus}
\begin{figure}[h]
  \centering
  \includegraphics[width=1\textwidth]{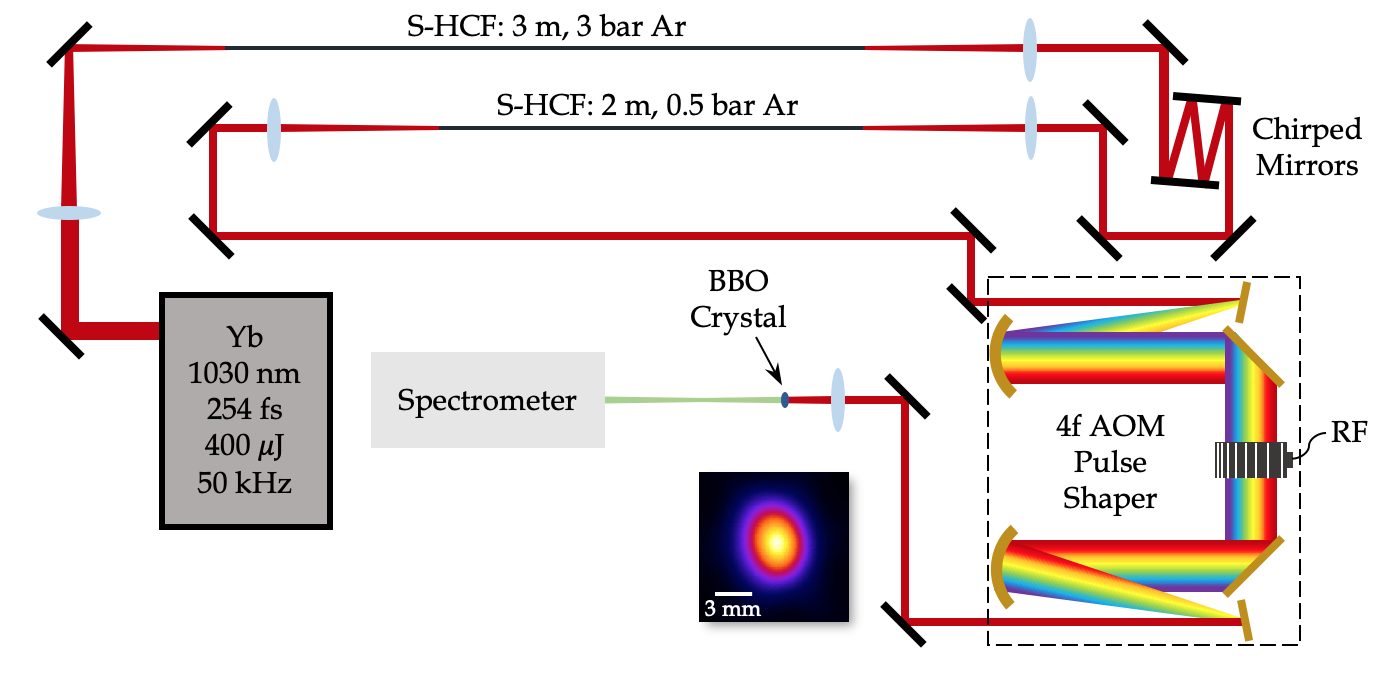}
  \caption{\label{Apparatus}Experimental Apparatus. The pulses from the Yb laser are centered at about 1025~nm, with a 250~fs duration. They are focused into a 3~m long HCF filled with 3~bar of Ar and compressed with chirped mirrors. They are then focused into a 2~m long HCF filled with 0.5~bar of Ar and sent into the AOM-based pulse-shaper (the inset shows the mode of the beam exiting the pulse-shaper). The compressed output of the pulse-shaper is sent through a Beta Barium Borate (BBO) crystal and into the spectrometer for characterization.}
  \label{fig:apparatus}
\end{figure}
 
The schematic in Figure~\ref{fig:apparatus} shows the two-stage pulse-compression system and the subsequent measurement apparatus. In brief, each stage consists of a spectral broadening component and a pulse compression component. Spectral broadening is achieved with Kerr-induced self-phase modulation by filling stretched, hollow-core fibers (S-HCF) with argon \cite{nisoli1996generation,nagy}. Pulse-compression is done in two ways: after the first fiber, there are chirped mirrors, and after the second fiber, there is an AOM-based pulse-shaper. 

\begin{figure}[h]
  \centering
  \includegraphics[width=.8\textwidth]{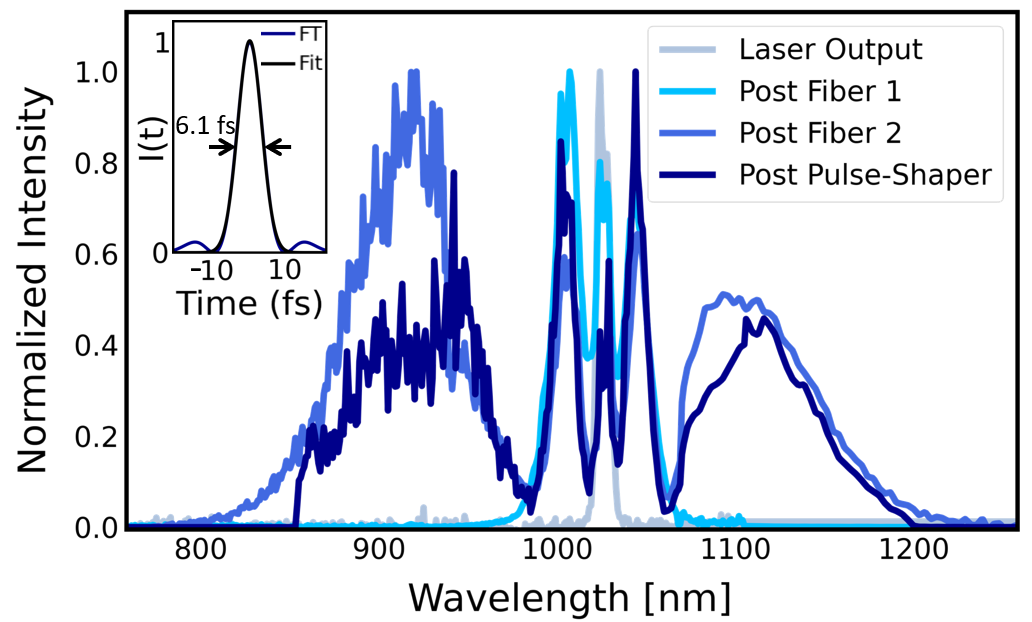}
  \caption{\label{Spectra}Initial spectrum out of the laser centered about 1025 nm (light blue), spectrum out of the first fiber with a bandwidth of 80 nm tail-to-tail (medium blue), spectrum out of the second fiber with a bandwidth of 400 nm (dark blue), and spectrum after the pulse-shaper with a bandwidth of 350 nm (gray). An inset of the transform limit after the second fiber is included.}
  \label{fig:spectra}
\end{figure}

Pulses from the Yb laser (50~kHz repetition rate, 400~$\mu$J pulse energy, 20~W average power, 250~fs pulse duration, 1025~nm central wavelength) are focused to a beam waist of 144~$\mu$m and sent into a 3~m S-HCF. The inner diameter of the fiber is 450~$\mu$m. The beam waist of 144~$\mu$m satisfies the mode matching condition \cite{abrams1972coupling}. The first fiber is glued onto mounts and stretched within a vacuum-sealable tube. The tube and enclosed fiber are filled with 3~bar of Ar. The transmission efficiency of the first fiber is 73\% including reflections from uncoated MgF$_2$ windows. With nano-structured or anti-reflection coated windows, an efficiency of 89\% should be possible. The output of the fiber is collimated, and the spectrally broadened pulses are compressed in duration with four bounces off -500~fs$^2$ chirped mirrors (Ultrafast Innovations HD59). We estimate the pulse duration to be approximately 35~fs after the first pulse-compression system based on calculations of the light—matter interaction in the fiber, and the Fourier transform limit of the spectrum (see Figure~2). The exit of the first fiber is imaged onto the entrance of a second 2~m long S-HCF with the same mode matching condition as the first. In comparison to the first fiber, the ends of the second are directly coupled to vacuum and movable to the beam path \cite{catanese2021acousto}; it is filled with 0.5 bar of Ar. The Fourier transform limit of the spectrum exiting the second fiber is 6~fs (Figure~\ref{fig:spectra}), and the efficiency is 80\%. The output of the second fiber is collimated and the pulse is compressed using an AOM-based pulse-shaper as opposed to chirped mirrors. 

The pulse-shaper is set up in a 4f configuration as discussed in earlier work \cite{jones2019dual, catanese2021acousto}, but optimized for a near-IR, high average power laser. Pulses with a bandwidth of 400~nm tail-to-tail are sent into the pulse-shaper and diffracted from a gold-coated ruled diffraction grating. The first order diffracted beam is propagated to a $f$ = 0.5~m, gold-coated curved mirror, which collimates the dispersed frequency components and focuses them into a TeO$_2$ AOM, which is placed in the Fourier plane of the zero dispersion stretcher geometry.  The combination of the grating and focusing mirror map the optical frequencies of the laser pulse to position in the AOM. Given the relatively slow velocity of the acoustic wave in the crystal, compared to the speed of light, the acoustic wave is seen as fixed with respect to the orthogonally propagating laser beam. Thus, the optical frequencies can be mapped to time in the acoustic wave. The maximum repetition rate of the laser is limited by the transit time of the acoustic wave through the active portion of the crystal. We utilize the entire 40~mm optical aperture to shape the optical bandwidth; this corresponds to 10~$\mu$s, or a maximum repetition rate of 100~kHz. However, the maximum laser output pulse energy at 100~kHz is half that at 50~kHz. In order to work with the maximum pulse energy from the laser, and avoid a 100\% duty cycle on the AOM, we work at 50~kHz. The shaped pulses exit the pulse-shaper in a symmetrically equivalent way to the input pulses. The efficiency of the 4f AOM-based pulse-shaper is about 25\%. 

In order to shape the pulse, we apply a mask with the acoustic wave onto the electric field of the pulse $E’(\omega) = M(\omega)E(\omega)$, where $E(\omega)$ is the input pulse, $E'(\omega)$ is the output shaped pulse, and $M(\omega)$ is the pulse-shaper applied mask function. The mask is applied to the pulse by controlling the amplitude and phase of the acoustic wave with the radio frequency voltage driving the modulator. While the mapping of acoustic time to optical wavelength is roughly linear with optical wavelength for small diffraction angles ($\sin(\theta)\sim \theta$), the mapping of acoustic time to optical frequency is highly nonlinear, given the inverse relationship between frequency and wavelength. This nonlinear mapping leads to a nonlinear optical phase as a function of frequency (higher order dispersion) \cite{jones2019dual} when programming a constant acoustic wave that diffracts the laser. However, depending on the orientation of the AOM and the acoustic diffraction order that one chooses, the sign of this dispersion can be chosen and utilized to compensate for normal dispersion from optics in the beam path \cite{catanese2021acousto}. Here, the transducer of the AOM is on the red side of the optical spectrum and the +1 diffraction order is used, which corresponds to a nonlinear mapping that introduces a  negative group delay dispersion. 

The mapping of optical frequency to position, as mentioned above, is important for applying the correct amount of dispersion to the pulse and for accurately decoupling the higher orders of dispersion. Dispersion is often described in terms of a Taylor series expansion. Thus, we utilize the same formulation for dispersion control of the pulse shaper and write the mask:
\begin{equation}
    M_{\text{phase}}(\omega) = \exp\bigg(i\frac{\text{GDD}}{2}(\omega-\omega_0)^2 + i\frac{\text{TOD}}{6}(\omega-\omega_0)^3 + i\frac{\text{FOD}}{24}(\omega-\omega_0)^4\bigg)
    \label{eq:compression}
\end{equation}
where $\omega_0$ is the central frequency of the expansion, GDD is the group-delay dispersion, TOD is the third-order dispersion, and FOD is the fourth-order dispersion. Two techniques were used to measure the pulse duration: D-scans and CFROGs. The pulses were sent through a 10~$\mu$m thick BBO crystal and the resulting second harmonic generated (SHG) spectrum was recorded. For D-scans, GDD was for fixed TOD and FOD by programming these values onto the pulse-shaper.

Similarly, for CFROGs, the dispersion (GDD, TOD, and FOD) is fixed, and the delay between pulses is programmed and scanned by the pulse-shaper utilizing the mask:
\begin{equation}
    M_{\text{CFROG}}(\omega) = 1+R\exp\big(i(\omega-\omega_L)\tau+i\phi_L\big)
    \label{eq:CFROG}
\end{equation}
where $R$ is the relative pulse amplitude, $\omega_L$ is the locking frequency (frequency at which constructive interference occurs independent of delay), $\tau$ is the pulse delay, and $\phi_L$ is the relative pulse phase. Since CFROGs use a collinear geometry, they are easier to set-up and remove the potential for temporal smearing associated with non-collinear FROG measurements. Moreover, normally the scan step would require greater than Nyquist sampling of the carrier, but an appropriate choice of $\omega_L$ softens this requirement such that one only needs a step size small enough to sample the envelope.

\section{Apparatus Discussion}
Multiple avenues of pulse compression were attempted to achieve sub-10~fs pulses. We originally tried to compress the 250~fs pulses out of the laser with a single S-HCF, filled with molecular gas, and the AOM-based pulse-shaper. We filled the fiber with molecular gas because spectral broadening can be enhanced by the rotational Raman process \cite{fan2020high}. The most promising gas seemed to be N$_2$O based on the Fourier transform limit of the spectrum out of the fiber. However, at average powers of 10~W or greater, the molecular gas absorbed too much energy from the laser as a result of rotational heating, dramatically decreasing the throughput efficiency and damaging the entrance to the optical fiber. Thus, we made use of atomic gases, for which the nonlinear process is strictly parametric, with no heating of the gas. This, however, required two fibers in order to produce the desired bandwidth and final pulse duration.  

High average power through the pulse-shaper also needed to be considered. Too high a peak intensity in the TeO$_2$ AOM can lead to self focusing and self phase modulation, and too high an average power on the diffraction gratings can lead to mode distortions.  While increasing the mode size into the pulse-shaper mitigates the mode distortions due to nonuniform thermal expansion of the gratings, it leads to a tighter focus in the AOM, which can lead to both self focusing and damage. Therefore, we had to strike a balance between these two requirements, and used a mode size of 2.5~mm FWHM going into the pulse-shaper. We explored two different pairs of gratings - one with 300~lines/mm, and another with 200~lines/mm. Both were gold-coated ruled diffraction gratings blazed for operation at about 1000~nm. They both consist of a glass substrate with an epoxy that was used to produce a replica grating from a ruled master, coated with a thin layer of chromium and then gold. Combined with the 0.5~m focal length mirrors, the dispersed spectrum with the 300~lines/mm grating was too large for the aperture of the AOM, limiting the spectrum and resulting in compression to only 20~fs. At average powers over 10~W, these gratings in the pulse-shaper became deformed by the thermal load from the laser and this decreased the efficiency of the pulse-shaper to 20\%, but the mode and power were stable. For the gratings with a groove density of 200~lines/mm, we were able to compress the pulses to $\sim$10~fs, but these gratings provided lower efficiency and were damaged at average powers over 10~W. We considered changing the curved mirrors to ones with a shorter focal length to fit more of the bandwidth on the AOM, but this would also produce a tighter focus in the AOM, leading to nonlinear phase accumulation in the TeO$_2$. In the future, we plan on improving the grating performance with a more appropriate substrate (copper, to avoid nonuniform thermal expansion) and a lower groove density (in order to fit the entire optical spectrum through the AOM). 
\begin{figure}[H]
  \centering
  \includegraphics[width=.7\textwidth]{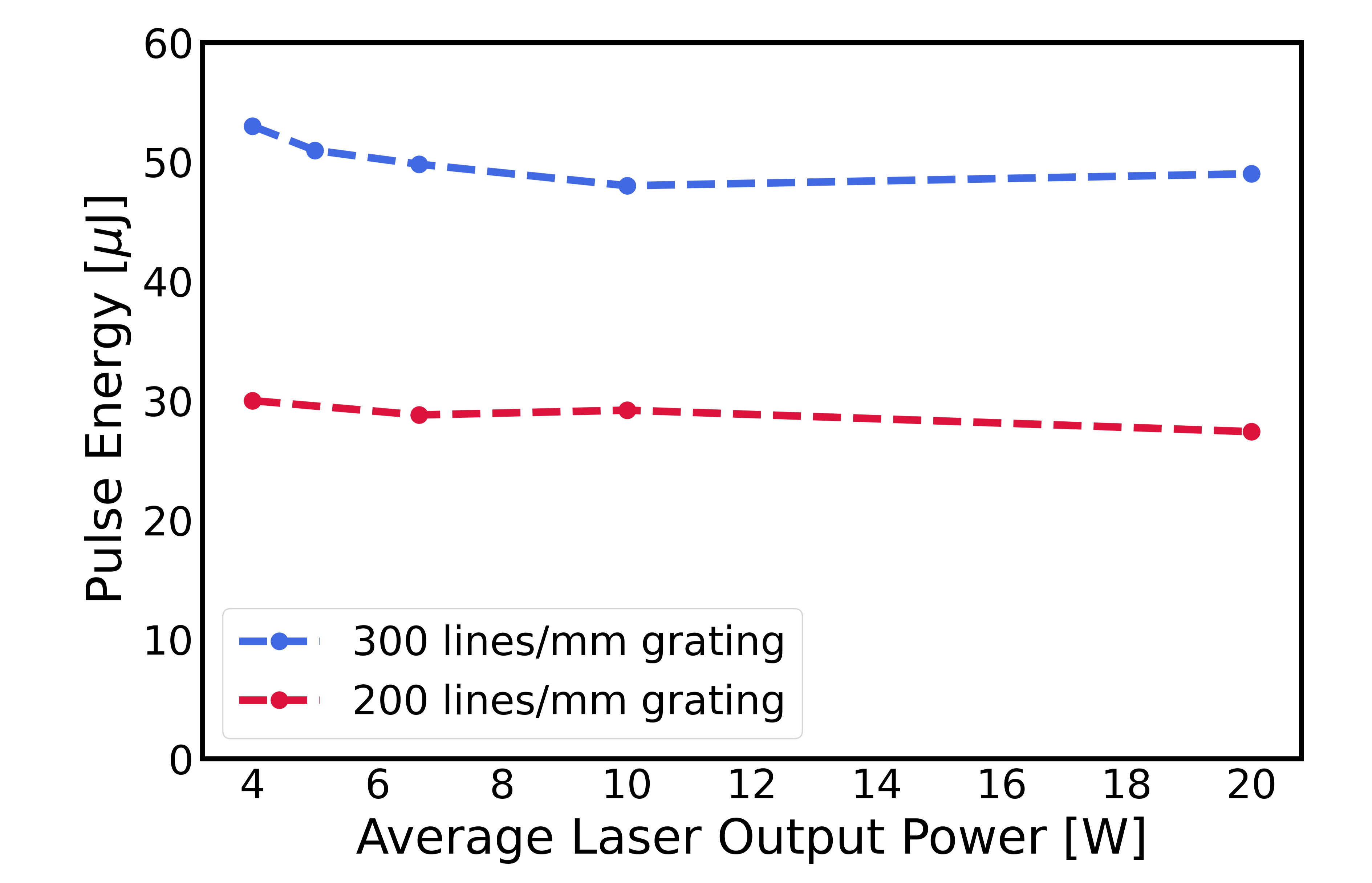}
  \caption{\label{Spectra} Overall pulse energy out of the system after the pulse-shaper with two different gratings: 300~lines/mm (blue) and 200~lines/mm (red).}
  \label{fig:outputEnergy}
\end{figure}

\section{Results}
\begin{figure}[H]
    \centering
    \includegraphics[width = 0.9\textwidth]{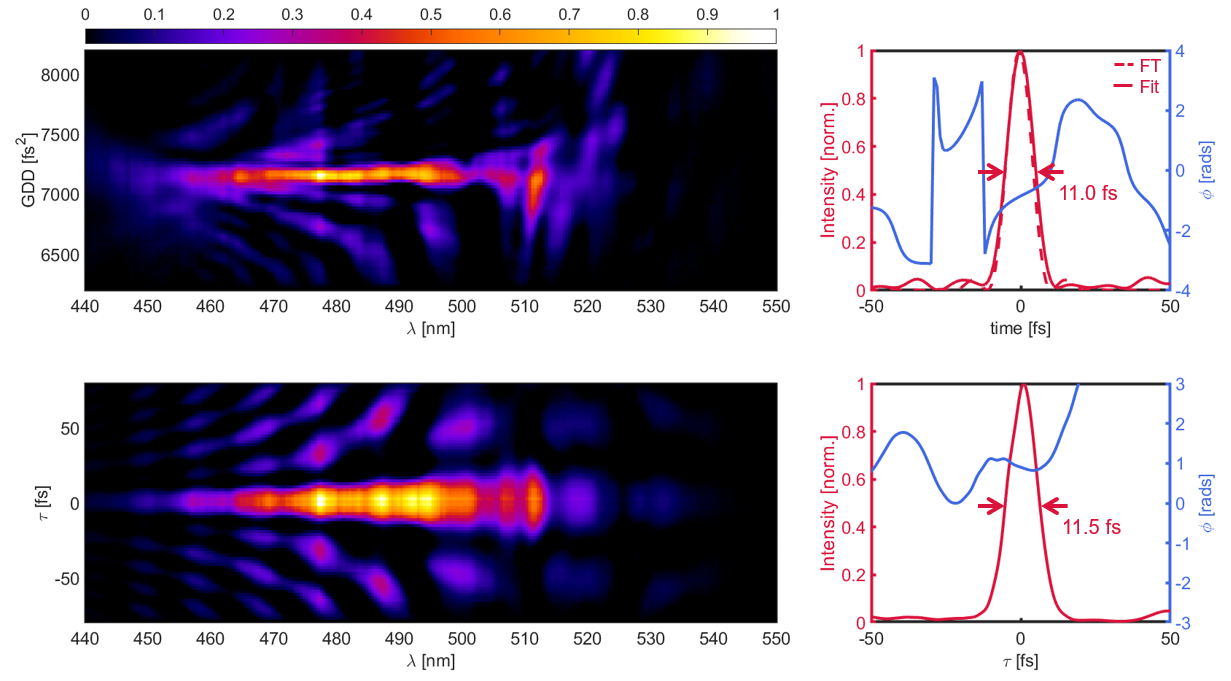}
    \caption{Top panel: an AOM-based pulse-shaper D-scan and its corresponding, reconstructed pulse duration (11.0 fs) and phase. Bottom panel: an AOM-based pulse-shaper CFROG and its corresponding, reconstructed pulse duration (11.5 fs) and phase.}
    \label{fig:dscancfrog}
\end{figure}

With this system, we are able to compress the 250~fs pulses out of the laser down to 11~fs. One of the strengths of the AOM-based pulse-shaper is that it allows for different types of pulse characterization. By writing a mask on the phase of the acoustic wave, we can scan through the GDD and adjust the TOD and FOD accordingly, i.e. Eq. \ref{eq:compression}. For CFROGs, we can utilize the same phase mask and scan the delay between two pulses (product of Eqs. \ref{eq:compression} and \ref{eq:CFROG}). Figure~\ref{fig:dscancfrog} shows a D-scan and CFROG that reconstruct to 11.0~fs and 11.5~fs, respectively.  We obtained shorter pulses in some cases ($<10$~fs), but with more structure in time.  In order to obtain the D-scan, the GDD was scanned from 6200 to 8000~fs$^2$, and the applied TOD and FOD were -23700~fs$^3$ and 55000~fs$^4$. With the acquired GDD, TOD, and FOD, we then scanned the delay between a pair of pulses with $\omega_L=1.86$ rad/fs and $\phi_L=0$. The reconstructed duration for both types of pulse characterization agree within 0.5~fs. The Fourier transform limited duration calculated with the bandwidth measured after the pulse-shaper was about 8~fs (Figure \ref{fig:spectra}), a longer duration than the spectra out of the fibers provide for due to optical clipping on the AOM crystal. We have included a table that compares our experimentally applied values of dispersion to the expected values we calculated. While we expected to achieve the shortest compressed pulse with an applied phase mask that cancels the sum of the elements (dispersive elements plus the phase associated with the nonlinear mapping of optical frequency to acoustic time), this was not exactly the case. The differences in the last row of the table are small but not insignificant. (If uncorrected, the residual GDD, TOD and FOD would individually lead to pulse durations of 120 fs, 23 fs and 21 fs respectively).  We attribute these differences mostly to small errors in the calibration of the optical frequency to acoustic time, with small contributions from our imperfect characterization of the phase accumulated in the spectral broadening and subsequent optical elements.  
%we note that the phase masks required for the shortest pulse yielded a The shortest compressed pulse should correspond to the sum of all the dispersive elements and the applied phase mask canceling each other out. However, the sum of these values does not equal 0 dispersion; the dispersion we have left over is 543~fs$^2$, -4635~fs$^3$, and 31835~fs$^4$. 
%The value for GDD is non-negligible, and would stretch out a 10~fs pulse to 150~fs. Yet, the TOD would only stretch the pulse to 23~fs and the FOD would stretch it to 21~fs. 
%Even though the values of TOD and FOD seem large, they have less of an effect on a short pulse than the GDD. We suspect our calculations for the expected dispersion or the accounting of dispersion in the beam path are not precise enough. 

We note that the compressed second harmonic spectra are not centered at 513~nm, which one might expect given the 1025~nm fundamental. The loss of the lower frequency components in the second harmonic spectra could be due to a number of reasons, including limited phase matching in the BBO crystal, frequency dependent losses in the pulse shaper, spatial chirp, etc.  We have ruled out most of these reasons and suspect that some spatial chirp combined with imperfect coupling into the spectrometer is the main contribution.  We are planning improvements to the collimation of the beam out of the fibers and pulse characterization to address this.
%Multiple potential reasons for this issue were checked and ruled out. The BBO crystal is 0.01~mm thick and calculations based on the thickness did not suggest a phase matching problem, nor did tuning the crystal angle change the spectra. Coupling into the spectrometer and a flaw in the program were eliminated as possible problems; the latter by doing a manual D-scan without the code. Our suspicion is that there is a residual spatial chirp that is not being compensated. 

\begin{table}[H]
\centering
\begin{tabular}{cccc} \toprule
 Dispersive Elements & GDD (fs$^2$) & TOD (fs$^3$) & FOD (fs$^4$) \\ \midrule %[0.5ex]
 Lens  & 68 & 123 & -136 \\
 %Window  & 61 & 121 & -141 \\
 %$\frac{1}{2}$ Waveplate  & 150 & 326 & -400 \\
 Air        & 65  & 9 & - \\
 AOM TeO$_2$  & 3618 & 2583 & 1172 \\
 Ar Broadening & 250 & - & - \\ 
 Nonlinear Mapping & -10624 & 16350 & -24201 \\ \midrule
 Sum of Elements & -6623 & 19065 & -23165 \\
 Applied Phase Mask & 7166 & -23700 & 55000 \\ \midrule\midrule
 Sum + Mask & 543 & -4635 & 31835 \\ \bottomrule
\end{tabular}
\caption{Table comparing the sum of the applied dispersion and dispersion accumulated along the table. To have the shortest compressed pulse, we expect the sum of the elements and the applied mask to cancel each other out.}
\label{tab:dtable}
\end{table}

\begin{figure}[H]
    \centering
    \includegraphics[width = .9\textwidth]{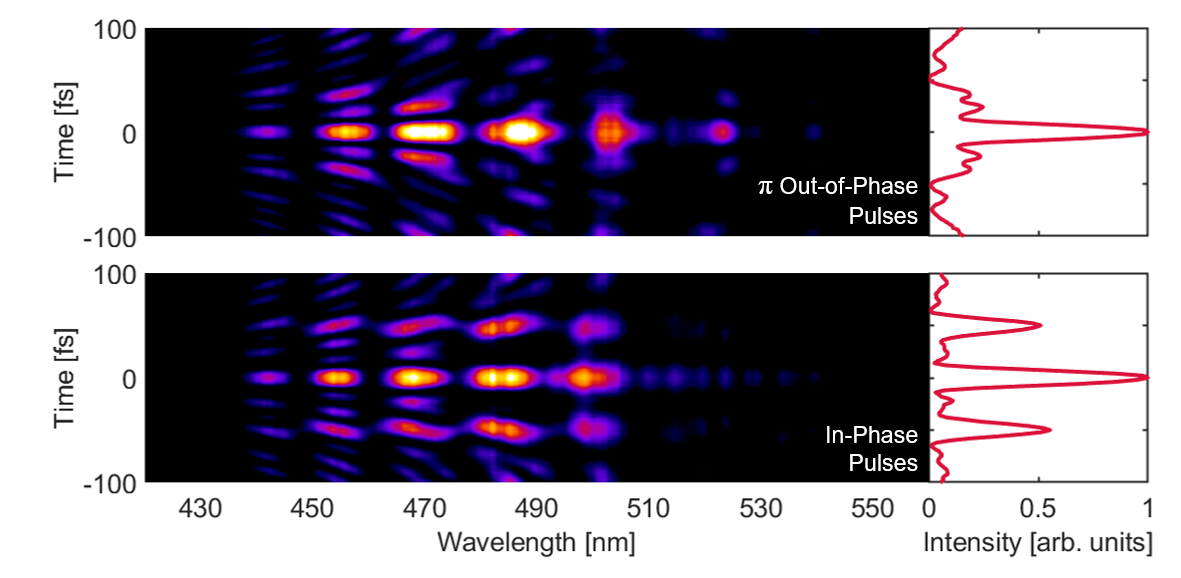}
    \caption{CFROG characterizations of double pulses where the pulses in the top panel were $\pi$ out-of-phase and the pulses in the bottom panel were in-phase. The plots to the right of the CFROGs are the integrated intensity of the CFROGs over wavelength.}
    \label{fig:doublePulse}
\end{figure}

Figure \ref{fig:doublePulse} is a testament to the ability of the pulse-shaper. It shows a CFROG characterization of two pulses that were delayed by $t=50$~fs, but the two pulses were $\pi$ out-of-phase with each other on the top panel and in-phase on the bottom panel. The effect of the different phases can be seen in the intensity plots to the right of the CFROGs. At 50~fs in the CFROG delay, the pulses destructively interfere when they are out-of-phase and constructively interfere when they are in-phase. For the double pulse, the mask is slightly different than the CFROG mask from above (Eq. \ref{eq:CFROG}). The CFROG mask is multiplied by a double pulse mask:
\begin{equation}
    M_\text{Double Pulse}(\omega) = \underbrace{\big[1+R\exp\big(i(\omega-\omega_L)\tau+i\phi_L\big)\big]}_\text{CFROG Delay}
    \times
    \underbrace{\big[1+\exp\big(i(\omega-\omega_L)t+i\phi\big)\big]}_\text{Double Pulse Delay}
    \label{eq:doublePulse}
\end{equation}
where $t$ is the delay between the double pulses and $\phi$ is the relative pulse phase between the double pulses. This mask (Eq. \ref{eq:doublePulse}) is also multiplied by the phase mask (Eq. \ref{eq:compression}) to compress the pulses. Independent control over the delay and phase can be very important for nonlinear spectroscopies, such as 2D electronic spectroscopy \cite{agathangelou2021phase, collini2010coherently}.

\section{Conclusion}
We have demonstrated the ability to achieve a high compression ratio of pulses with high pulse energy and repetition rate. The pulses can be shaped in phase and amplitude with the AOM-based pulse-shaper, and pulse pairs can be created with independent control over delay and phase. This allows us to both create ultrafast pulses, on the order of 10~fs, and characterize them with the same device. Currently, we are limited by the efficiency and groove density of our diffraction gratings, but have shown the capability of ruled diffraction gratings to handle the high average power and large bandwidths. The generation of intense, ultrafast pulses with arbitrary delays between pulse pairs will allow for pump-probe measurements of molecular dynamics \cite{tagliamonti2017time}.

\section*{Funding}
National Science Foundation (2110376).
\section*{Acknowledgments}
We thank Joseph Robinson (SLAC) and John Travers (Heriot Watt) for valuable discussion as well as Light Conversion for supporting this work. This work was funded by the Linac Coherent Light Source, SLAC National Accelerator Laboratory, which is supported by the U.S. Department of Energy, Office of Science, Office of Basic Energy Sciences, under Contract No. DE-AC02-76SF00515 and the National Science Foundation under award number 2110376.
\section*{Disclosures} 
The authors declare no conflicts of interest.
\section*{Data availability}
Data underlying the results presented in this paper are not publicly available at this time but may
be obtained from the authors upon reasonable request.

\bibliography{Biblio}%% use BibTeX or add references manually

\begin{thebibliography}{10}
\newcommand{\enquote}[1]{``#1''}

\bibitem{ullrich2003recoil}
J.~Ullrich, R.~Moshammer, A.~Dorn, R.~D{\"o}rner, L.~P.~H. Schmidt, and
  H.~Schmidt-B{\"o}cking, \enquote{Recoil-ion and electron momentum
  spectroscopy: reaction-microscopes,} {\protect\JournalTitle{Reports on
  progress in physics}} \textbf{66}, 1463 (2003).

\bibitem{zhao2017coincidence}
A.~Zhao, M.~van Beuzekom, B.~Bouwens, D.~Byelov, I.~Chakaberia, C.~Cheng,
  E.~Maddox, A.~Nomerotski, P.~Svihra, J.~Visser \emph{et~al.},
  \enquote{Coincidence velocity map imaging using tpx3cam, a time stamping
  optical camera with 1.5 ns timing resolution,} {\protect\JournalTitle{Review
  of Scientific Instruments}} \textbf{88} (2017).

\bibitem{boguslavskiy2012multielectron}
A.~E. Boguslavskiy, J.~Mikosch, A.~Gijsbertsen, M.~Spanner, S.~Patchkovskii,
  N.~Gador, M.~J. Vrakking, and A.~Stolow, \enquote{The multielectron
  ionization dynamics underlying attosecond strong-field spectroscopies,}
  {\protect\JournalTitle{Science}} \textbf{335}, 1336--1340 (2012).

\bibitem{goulielmakis2008single}
E.~Goulielmakis, M.~Schultze, M.~Hofstetter, V.~S. Yakovlev, J.~Gagnon,
  M.~Uiberacker, A.~L. Aquila, E.~Gullikson, D.~T. Attwood, R.~Kienberger
  \emph{et~al.}, \enquote{Single-cycle nonlinear optics,}
  {\protect\JournalTitle{Science}} \textbf{320}, 1614--1617 (2008).

\bibitem{sandor2016strong}
P.~S{\'a}ndor, V.~Tagliamonti, A.~Zhao, T.~Rozgonyi, M.~Ruckenbauer,
  P.~Marquetand, and T.~Weinacht, \enquote{Strong field molecular ionization in
  the impulsive limit: Freezing vibrations with short pulses,}
  {\protect\JournalTitle{Physical review letters}} \textbf{116}, 063002 (2016).

\bibitem{zherebtsov2011controlled}
S.~Zherebtsov, T.~Fennel, J.~Plenge, E.~Antonsson, I.~Znakovskaya, A.~Wirth,
  O.~Herrwerth, F.~S{\"u}{\ss}mann, C.~Peltz, I.~Ahmad \emph{et~al.},
  \enquote{Controlled near-field enhanced electron acceleration from dielectric
  nanospheres with intense few-cycle laser fields,}
  {\protect\JournalTitle{Nature Physics}} \textbf{7}, 656--662 (2011).

\bibitem{fetterman1998ultrafast}
M.~R. Fetterman, D.~Goswami, D.~Keusters, W.~Yang, J.-K. Rhee, and W.~S.
  Warren, \enquote{Ultrafast pulse shaping: amplification and
  characterization,} {\protect\JournalTitle{Optics Express}} \textbf{3},
  366--375 (1998).

\bibitem{nisoli1996generation}
M.~Nisoli, S.~De~Silvestri, and O.~Svelto, \enquote{Generation of high energy
  10 fs pulses by a new pulse compression technique,}
  {\protect\JournalTitle{Applied Physics Letters}} \textbf{68}, 2793--2795
  (1996).

\bibitem{nagy2020generation}
T.~Nagy, M.~Kretschmar, M.~J. Vrakking, and A.~Rouz{\'e}e, \enquote{Generation
  of above-terawatt 1.5-cycle visible pulses at 1 khz by post-compression in a
  hollow fiber,} {\protect\JournalTitle{Optics Letters}} \textbf{45},
  3313--3316 (2020).

\bibitem{silletti2023dispersion}
L.~Silletti, A.~Bin~Wahid, E.~Escoto, P.~Balla, S.~Rajhans, K.~Horn,
  L.~Winkelmann, V.~Wanie, A.~Trabattoni, C.~M. Heyl \emph{et~al.},
  \enquote{Dispersion-engineered multi-pass cell for single-stage
  post-compression of an ytterbium laser,} {\protect\JournalTitle{Optics
  Letters}} \textbf{48}, 1842--1845 (2023).

\bibitem{lu2018sub}
C.-H. Lu, T.~Witting, A.~Husakou, M.~J. Vrakking, A.~Kung, and F.~J. Furch,
  \enquote{Sub-4 fs laser pulses at high average power and high repetition rate
  from an all-solid-state setup,} {\protect\JournalTitle{Optics express}}
  \textbf{26}, 8941--8956 (2018).

\bibitem{he2017high}
P.~He, Y.~Liu, K.~Zhao, H.~Teng, X.~He, P.~Huang, H.~Huang, S.~Zhong, Y.~Jiang,
  S.~Fang, X.~Hou, and Z.~Wei, \enquote{High-efficiency supercontinuum
  generation in solid thin plates at 0.1 {TW} level,}
  {\protect\JournalTitle{Opt. Lett.}} \textbf{42}, 474--477 (2017).

\bibitem{hauri2004generation}
C.~P. Hauri, W.~Kornelis, F.~W. Helbing, A.~Heinrich, A.~Couairon,
  A.~Mysyrowicz, J.~Biegert, and U.~Keller, \enquote{Generation of intense,
  carrier-envelope phase-locked few-cycle laser pulses through filamentation,}
  {\protect\JournalTitle{Appl. Phys. B}} \textbf{79}, 673--677 (2004).

\bibitem{nagy2021high}
T.~Nagy, P.~Simon, and L.~Veisz, \enquote{High-energy few-cycle pulses:
  post-compression techniques,} {\protect\JournalTitle{Adv. Phys.}} \textbf{6},
  1845795 (2021).

\bibitem{Gallmann2007Comparison}
L.~Gallmann, T.~Pfeifer, P.~Nagel, M.~Abel, D.~Neumark, and S.~Leone,
  \enquote{Comparison of the filamentation and the hollow-core fiber
  characteristics for pulse compression into the few-cycle regime,}
  {\protect\JournalTitle{Appl. Phys. B}} \textbf{86}, 561–566 (2007).

\bibitem{mayer1997ultrabroadband}
E.~Mayer, J.~M{\"o}bius, A.~Euteneuer, W.~R{\"u}hle, and R.~Szip{\H{o}}cs,
  \enquote{Ultrabroadband chirped mirrors for femtosecond lasers,}
  {\protect\JournalTitle{Optics letters}} \textbf{22}, 528--530 (1997).

\bibitem{nagy2019generation}
T.~Nagy, S.~H{\"a}drich, P.~Simon, A.~Blumenstein, N.~Walther, R.~Klas,
  J.~Buldt, H.~Stark, S.~Breitkopf, P.~J{\'o}j{\'a}rt \emph{et~al.},
  \enquote{Generation of three-cycle multi-millijoule laser pulses at 318 w
  average power,} {\protect\JournalTitle{Optica}} \textbf{6}, 1423--1424
  (2019).

\bibitem{cardin20150}
V.~Cardin, N.~Thir{\'e}, S.~Beaulieu, V.~Wanie, F.~L{\'e}gar{\'e}, and B.~E.
  Schmidt, \enquote{0.42 tw 2-cycle pulses at 1.8 $\mu$ m via hollow-core fiber
  compression,} {\protect\JournalTitle{Applied Physics Letters}} \textbf{107},
  181101 (2015).

\bibitem{gebhardt2017nonlinear}
M.~Gebhardt, C.~Gaida, T.~Heuermann, F.~Stutzki, C.~Jauregui, J.~Antonio-Lopez,
  A.~Schulzgen, R.~Amezcua-Correa, J.~Limpert, and A.~T{\"u}nnermann,
  \enquote{Nonlinear pulse compression to 43 w gw-class few-cycle pulses at 2
  $\mu$m wavelength,} {\protect\JournalTitle{Optics letters}} \textbf{42},
  4179--4182 (2017).

\bibitem{timmers2017generating}
H.~Timmers, Y.~Kobayashi, K.~F. Chang, M.~Reduzzi, D.~M. Neumark, and S.~R.
  Leone, \enquote{Generating high-contrast, near single-cycle waveforms with
  third-order dispersion compensation,} {\protect\JournalTitle{Optics letters}}
  \textbf{42}, 811--814 (2017).

\bibitem{jones2019dual}
A.~C. Jones, M.~B. Kunz, I.~Tigges-Green, and M.~T. Zanni, \enquote{Dual
  spectral phase and diffraction angle compensation of a broadband aom 4-f
  pulse-shaper for ultrafast spectroscopy,} {\protect\JournalTitle{Optics
  express}} \textbf{27}, 37236--37247 (2019).

\bibitem{catanese2021acousto}
A.~Catanese, B.~Kaufman, C.~Cheng, E.~Jones, M.~G. Cohen, and T.~Weinacht,
  \enquote{Acousto-optic modulator pulse-shaper compression of octave-spanning
  pulses from a stretched hollow-core fiber,} {\protect\JournalTitle{OSA
  Continuum}} \textbf{4}, 3176--3183 (2021).

\bibitem{miranda2012characterization}
M.~Miranda, C.~L. Arnold, T.~Fordell, F.~Silva, B.~Alonso, R.~Weigand,
  A.~L’Huillier, and H.~Crespo, \enquote{Characterization of broadband
  few-cycle laser pulses with the d-scan technique,}
  {\protect\JournalTitle{Optics express}} \textbf{20}, 18732--18743 (2012).

\bibitem{amat2004ultrashort}
I.~Amat-Rold{\'a}n, I.~G. Cormack, P.~Loza-Alvarez, E.~J. Gualda, and
  D.~Artigas, \enquote{Ultrashort pulse characterisation with shg
  collinear-frog,} {\protect\JournalTitle{Optics express}} \textbf{12},
  1169--1178 (2004).

\bibitem{nagy}
T.~Nagy, P.~Simon, and L.~Veisz, \enquote{High-energy few-cycle pulses:
  post-compression techniques,} {\protect\JournalTitle{Advances in Physics: X}}
  \textbf{6}, 1845795 (2021).

\bibitem{abrams1972coupling}
R.~Abrams, \enquote{Coupling losses in hollow waveguide laser resonators,}
  {\protect\JournalTitle{IEEE Journal of Quantum Electronics}} \textbf{8},
  838--843 (1972).

\bibitem{fan2020high}
G.~Fan, R.~Safaei, O.~Kwon, V.~Schuster, K.~L{\'e}gar{\'e}, P.~Lassonde,
  A.~Ehteshami, L.~Arias, A.~Laram{\'e}e, J.~Beaudoin-Bertrand \emph{et~al.},
  \enquote{High energy redshifted and enhanced spectral broadening by molecular
  alignment,} {\protect\JournalTitle{Optics Letters}} \textbf{45}, 3013--3016
  (2020).

\bibitem{agathangelou2021phase}
D.~Agathangelou, A.~Javed, F.~Sessa, X.~Solinas, M.~Joffre, and J.~P. Ogilvie,
  \enquote{Phase-modulated rapid-scanning fluorescence-detected two-dimensional
  electronic spectroscopy,} {\protect\JournalTitle{The Journal of Chemical
  Physics}} \textbf{155} (2021).

\bibitem{collini2010coherently}
E.~Collini, C.~Y. Wong, K.~E. Wilk, P.~M. Curmi, P.~Brumer, and G.~D. Scholes,
  \enquote{Coherently wired light-harvesting in photosynthetic marine algae at
  ambient temperature,} {\protect\JournalTitle{Nature}} \textbf{463}, 644--647
  (2010).

\bibitem{tagliamonti2017time}
V.~Tagliamonti, B.~Kaufman, A.~Zhao, T.~Rozgonyi, P.~Marquetand, and
  T.~Weinacht, \enquote{Time-resolved measurement of internal conversion
  dynamics in strong-field molecular ionization,}
  {\protect\JournalTitle{Physical Review A}} \textbf{96}, 021401 (2017).

\end{thebibliography}
\end{document}